\documentclass[11pt,a4paper]{article}
\usepackage{authblk}
\usepackage{amsmath,amsfonts,amssymb}
\usepackage{bbold}
\usepackage{bbm}
\usepackage{txfonts}
\usepackage{tensor}
\usepackage[T1]{fontenc}
\usepackage{hyperref}

\newtheorem{proof}{Proof}

\def\PL#1{\textit{Phys.\ Lett.}\ {\bf#1}}\def\CMP#1{Commun.\ Math.\ Phys.\ {\bf#1}}

\def\PR#1{\textit{Phys.\ Rev.}\ {\bf#1}}
\def\NP#1{\textit{Nucl.\ Phys.}\ {\bf#1}}
\def\JMP#1{\textit{J.\ Math.\ Phys.}\ {\bf#1}}

\def\JoP#1{\textit{J.\ Phys.}\ {\bf#1}} \def\IJMP#1{\textit{Int.\ J. Mod.\ Phys.}\ {\bf #1}}
\def\MPL#1{\textit{Mod.\ Phys.\ Lett.}\ {\bf #1}} \def\EL#1{\textit{Europhys.\ Lett.}\ {\bf #1}}

\def\AoP#1{\textit{Ann.\ Phys.}\ {\bf#1}}

\def\EPJ#1{\textit{Eur.\ Phys.\ J.}\ {\bf#1}}

\newcommand{\bibp}[3]{#1, #2, #3}

\title{Generalized Heisenberg algebra from $o(2,4)$ Yang model\footnote{ Dedicated to the memory of Professor Jerzy Lukierski}}

\begin{document}
\author[a]{Tea Martini\'{c} Bila\'{c}}
\affil[a]{Faculty of Science, University of Split, Rudjera Bo\v{s}kovi\'{c}a 33, 21000 Split, Croatia\\
	E-mail:\;\href{mailto:teamar@pmfst.hr}{teamar@pmfst.hr}}

\author[b]{Stjepan Meljanac}
\affil[b]{Rudjer Bo\v{s}kovi\'{c} Institute, Theoretical Physics Division, Bijeni\v{c}ka c. 54, HR 10002 Zagreb, Croatia\\
	E-mail:\;\href{mailto:meljanac@irb.hr}{meljanac@irb.hr}}

 \author[c]{Salvatore Mignemi}
 \affil[c]{Dipartimento di Matematica, Universit\`{a} di Cagliari via Ospedale 72, 09124 Cagliari, Italy and INFN, Sezione di Cagliari Cittadella Universitaria, 09042 Monserrato, Italy\\
 	E-mail:\;\href{mailto:smignemi@unica.it}{smignemi@unica.it}}

\date{}
\maketitle

\begin{abstract}
It is well known that the algebra $o(2,4)$ generates the conformal group, but it can also be used to define some variants of the Yang model of noncommutative geometry on a curved spacetime.
Starting from these examples, we construct a new physical model based on $o(2,4)$, that can be interpreted as a generalization of the Heisenberg algebra n phase space, with flat positions and momenta, but  nontrivial commutation relations between positions and momenta and with the Planck constant promoted to an operator.
\end{abstract}

\section {Introduction}
A given group and its related Lie algebra are often used to describe the symmetries of different physical models. For example, the hydrogen atom enjoys the same SO$(4)$ symmetry as a $3$-sphere. Another example is given by the SO$(2, 4)$ group. It is well known that the $o(2, 4)$ algebra generates the conformal group. However also the Yang model of noncommutative geometry on a curved background is based on the same algebra, for some values of its parameters. Clearly the physical interpretation of the two models is different, but one can use known representations of one of the algebras to get new representations of the other. The isomorphism between the Yang and conformal algebras was originally noticed in \cite{Ka2} and more recently discussed in \cite{Na} in the context of representation theory.

We recall that the conformal algebra is an extension of the Poincar\'{e} algebra that describes the symmetries of spacetime under dilatations and conformal transformations and is the spacetime  symmetry associated to massless particles \cite{Ka,bar}. The Yang model is instead a quantum phase space with noncommutative positions and momenta \cite{Ya}. The noncommutativity of position operators can be associated to a noncommutative geometry and then to a curved momentum space. Analogously, noncommutativity of momenta is usually associated to curved spacetimes.

The Yang model depends on two parameters, $\alpha$ and $\beta$ proportional to $1/R^{2}$ and to $1/M^{2}$, respectively, where $M$ is a mass parameter of the order of Planck mass and $R$ is a length parameter, usually related to the cosmological constant. When $\alpha$ and $\beta$ have opposite sign, the Yang algebra is isomorphic to the $o(2, 4)$ algebra, which describes either a Snyder space on an anti-de Sitter background or an anti-Snyder space on a de Sitter background. In recent years several aspects of the theory and some generalizations  have been investigated \cite{chr}-\cite{kup}.

In this letter, we give another interpretation of $o(2,4)$. Starting from the Yang $o(2,4)$ model, we construct two generalizations of the Heisenberg algebra that depend on one parameter $1/MR$, and whose commutation relations are identical to those of the conformal algebra for $MR = 1$. The Planck constant of the Heisenberg algebra  is generalized to an operator that corresponds to the dilatation generator in the conformal algebra. The possibility of a
variable Planck 'constant' has been previously considered in several papers, from different points of view, see for example refs.~\cite{CG}.

The physical interpretations of the generalized Heisenberg algebra, the Yang model and the conformal algebra are different and relations between them are presented.

\section{Generalized Heisenberg algebras}

Let us start with the Yang algebra isomorphic to $o(2,4)$ \cite{Ya,MM-2}
\begin{equation}\label{1.1}
\left[  \hat{x}_{\mu},\hat{x}_{\nu}\right]  =\frac{i\epsilon}{M^{2}}M_{\mu\nu}, \quad \left[  \hat{p}_{\mu},\hat{p}_{\nu}\right] =\frac{-i\epsilon}{R^{2}}M_{\mu\nu},
\end{equation}
\begin{equation}\label{1.2}
\left[  M_{\mu\nu}, \hat{x}_{\lambda}\right]   =i\left( \eta_{\mu\lambda}\hat{x}_{\nu}-\eta_{\nu\lambda}\hat{x}_{\mu}\right) ,\quad  \left[  M_{\mu\nu}, \hat{p}_{\lambda}\right]   =i\left( \eta_{\mu\lambda}\hat{p}_{\nu}-\eta_{\nu\lambda}\hat{p}_{\mu}\right),
\end{equation}
\begin{equation}\label{1.3}
\left[  \hat{x}_{\mu}, \hat{p}_{\nu}\right] =i\eta_{\mu\nu}\hat{h}, \quad \left[  \hat{h}, \hat{x}_{\mu}\right]  =\frac{i\epsilon}{M^{2}}\hat{p}_{\mu}, \quad \left[  \hat{h}, \hat{p}_{\mu}\right]  =-\frac{i\epsilon}{R^{2}}\hat{x}_{\mu},
	\end{equation}
\begin{equation}\label{1.4}
\left[ M_{\mu\nu}, \hat{h} \right]  =0,
\end{equation}
\begin{equation}\label{1.5}
\left[  M_{\mu\nu},M_{\rho\sigma}\right]  =i\left( \eta_{\mu\rho}M_{\nu\sigma}-\eta_{\mu\sigma}M_{\nu\rho}-\eta_{\nu\rho}M_{\mu\sigma}+\eta_{\nu\sigma}M_{\mu\rho}\right),
\end{equation}
where $\epsilon^{2}=1$ and all generators are Hermitian. The operators $\hat x_\mu$ and $\hat p_\mu$ are interpreted as phase space coordinates, while the $M_{\mu\nu}$ generate the Lorentz algebra $o(1,3)$.

We consider a special case of  new generators linear in $\hat{x}_{\mu},\; \hat{p}_{\mu}\;$ and $M_{\mu\nu}$,
\begin{equation}\label{x-2}
\tilde{X}_{\mu}=\frac{1}{\sqrt{2}}\left(  \hat{x}_{\mu}+\frac{R}{M} \hat{p}_{\mu}\right) , \quad \tilde{P}_{\mu}= \frac{1}{\sqrt{2}}\left( \hat{p}_{\mu}-\frac{M}{R}\hat{x}_{\mu}\right) ,
\end{equation}
with inverse transformations  given by
\begin{equation}\label{1.6}
\hat{x}_{\mu}=\frac{1}{\sqrt{2}}\left( \tilde{X}_{\mu}-\frac{R}{M}\tilde{P}_{\mu}\right), \quad \hat{p}_{\mu}=\frac{1}{\sqrt{2}}\left( \tilde{P}_{\mu}+\frac{M}{R}\tilde{X}_{\mu}\right) .
\end{equation}

The generators $\tilde{X}_{\mu}, \;\tilde{P}_{\mu},\; M_{\mu\nu}$ and $\tilde{H}$ generate a new class of Lie algebras isomorphic to the initial Yang algebra. These algebras are defined by
\begin{equation}\label{1.8}
\left[  \tilde{X}_{\mu}, \tilde{X}_{\nu} \right]  =
\left[  \tilde{P}_{\mu}, \tilde{P}_{\nu}\right]  =0,
\end{equation}
\begin{equation}\label{1.9}
\left[  \tilde{X}_{\mu}, \tilde{P}_{\nu}\right] =i\left( \eta_{\mu\nu}\tilde{H}-\frac{\epsilon}{MR} M_{\mu\nu}\right) ,
\end{equation}
\begin{equation}\label{1.10}
\left[  M_{\mu\nu},\tilde{X}_{\lambda}\right]  =i\left( \eta_{\mu\lambda}\tilde{X}_{\nu}-\eta_{\nu\lambda}\tilde{X}_{\mu} \right),
\end{equation}
\begin{equation}\label{1.11}
\left[  M_{\mu\nu},\tilde{P}_{\lambda}\right]  =i\left( \eta_{\mu\lambda}\tilde{P}_{\nu}-\eta_{\nu\lambda}\tilde{P}_{\mu} \right),
\end{equation}
where $\tilde{H}=\hat{h}$ and
\begin{equation}\label{1.12}
\left[  M_{\mu\nu},\tilde{H}\right] =0, \quad
\left[  \tilde{H}, \tilde{X}_{\mu}\right]  =\frac{i \epsilon}{MR} \tilde{X}_{\mu}, \quad
\left[  \tilde{H}, \tilde{P}_{\mu}\right] =\frac{-i\epsilon}{MR} \tilde{P}_{\mu}.
\end{equation}
We interpret $ \tilde{X}_{\mu}$ and $ \tilde{P}_{\mu}$ as the coordinates of a new flat phase space, with $ \tilde{H}$ a operatorial generalization of the Planck constant,
and we call this algebra generalized Heisenberg algebra. In fact, it can be considered as a deformation of the Heisenberg-Lorentz algebra by a parameter ${1\over MR}$, which reduces to the standard algebra when ${1\over MR}\to0$.
 For $\epsilon=-1$ and $MR=1$ the commutation relations are identical to those of the conformal algebra, reported in sect.~4.

Inspired by the standard realization of the conformal algebra \cite{Ka}, we construct new Hermitian realizations of the generators $\tilde{X}_{\mu}$ and $\tilde{P}_{\nu}$
in terms of canonical phase space operators $x_{\mu}$ and $p_{\nu}$ that satisfy the Heisenberg algebra
\begin{equation}\label{2}
	\left[ x_{\mu}, x_{\nu}\right] =\left[ p_{\mu}, p_{\nu}\right] =0, \quad \left[ x_{\mu}, p_{\nu}\right] =i \eta_{\mu\nu}.
\end{equation}
Then
\begin{equation}
	M_{\mu\nu}=x_{\mu}p_{\nu}-x_{\nu}p_{\mu},
\end{equation}
\begin{equation}\label{1.15}
	\tilde{X}_{\mu}= x_{\mu}-\frac{\epsilon}{2MR}\left( x_{\mu}x\cdot p -\frac{1}{2} x^{2}p_{\mu} +h.c. \right),
\end{equation}
\begin{equation}\label{1.16}
	\tilde P_{\mu}=p_{\mu},
\end{equation}
satisfying \eqref{1.8}. In analogy with \cite{MM}, eq. \eqref{1.9} is then satisfied by choosing the "Planck operator" $\tilde H$ as
\begin{align}
	\tilde{H}&=1-\frac{\epsilon}{2MR}\left( x\cdot p+p\cdot x\right) \label{1.17}\\
	&=\sqrt{1-\frac{\epsilon}{MR}\left( \tilde{X}\cdot\tilde{P}+\tilde{P}\cdot\tilde{X}\right) +\frac{1}{2M^{2}R^{2}}M_{\mu\nu}M^{\mu\nu}+\frac{4}{M^{2}R^{2}}}.\label{1.18}
\end{align}
Note that in this representation
\begin{equation}	
\hat{h}=\sqrt{1+\frac{\epsilon}{R^{2}}\hat{x}^{2}-\frac{\epsilon}{M^{2}}\hat{p}^{2}+\frac{1}{2M^{2}R^{2}}M_{\mu\nu}M^{\mu\nu}+\frac{4}{M^{2}R^{2}}}.
\end{equation}
The last expression differs from the one of ref.~\cite{MM} by last term $ \frac{4}{M^{2}R^{2}}$.

Defining the action of the operators as
\begin{equation}
	P_\mu\rhd1=0,\qquad\tilde f(\tilde X)\rhd1=f(x),\qquad \tilde g(\tilde X)\rhd1=g(x),
\end{equation}
one has
\begin{equation}
f(x)\star g(x)=\tilde f(\tilde X)\tilde g(\tilde X)\rhd1=\tilde g(\tilde X)\tilde f(\tilde X)\rhd1=g(x)\star f(x)\neq f(x)g(x).
\end{equation}
Hence, the star product related to the generalized Heisenberg algebra is commutative, but not pointwise.

In fact,
\begin{equation}
\tilde X_\mu\rhd1=x_\mu\left(1+\frac{2i\epsilon}{MR}\right),
\end{equation}
and
\begin{equation}
\tilde X_\mu\tilde X_\nu\rhd1=\left(1+\frac{2i\epsilon}{MR}\right)\left[x_\mu x_\nu+\frac{3i\epsilon}{MR}x_\mu x_\nu-\frac{i\epsilon}{2MR}\eta_{\mu\nu}x^2\right].
\end{equation}
It follows that
\begin{equation}
x_\mu\star x_\nu={1\over1+\frac{2i\epsilon}{MR}}\left[\left(1+\frac{3i\epsilon}{MR}\right)x_\mu x_\nu-\frac{i\epsilon}{2MR}\eta_{\mu\nu}x^2\right].
\end{equation}

Moreover, the coproduct is nontrivial,
\begin{equation}
\Delta p_\mu\neq p_\mu\otimes 1+1\otimes p_\mu.
\end{equation}
More details on the star product, the coproduct and the Hopf algebra will be presented elsewhere.

\section{Yang algebra}
We also propose an exact realization of the  Yang algebra \eqref{1.1}-\eqref{1.5} alternative to those presented in ref.~\cite{MM-2}, which can be written  in terms of new variables
\begin{equation}
\xi_{\mu}=\frac{1}{\sqrt{2}}\left( x_{\mu}-\frac{R}{M}p_{\mu}\right), \quad
\pi_{\mu}=\frac{1}{\sqrt{2}}\left( p_{\mu}+\frac{M}{R}x_{\mu}\right),
\end{equation}
with inverse
\begin{equation}
x_{\mu}=\frac{1}{\sqrt{2}}\left( \xi_{\mu}+\frac{R}{M}\pi_{\mu}\right), \quad
p_{\mu}=\frac{1}{\sqrt{2}}\left( \pi_{\mu}-\frac{M}{R}\xi_{\mu}\right),
\end{equation}
that satisfy canonical commutation relations,
\begin{equation}
	\left[\xi_{\mu}, \xi_{\nu}\right] =\left[\pi_{\mu}, \pi_{\nu}\right] =0, \quad \left[\xi_{\mu},\pi_{\nu}\right] =i \eta_{\mu\nu}.
\end{equation}

 The simplest realization of this kind   is
\begin{equation}\label{2.1}
\hat{x}_{\mu}=\xi_{\mu}+\frac{\epsilon}{2}\left( \xi_{\mu}\left( -\frac{3}{8} u + \frac{1}{8} v- \frac{1}{4}z\right)+\frac{R}{M} \left(-\frac{1}{8} u + \frac{3}{8} v+ \frac{1}{4}z\right)\pi_{\mu} + h.c. \right),
\end{equation}
\begin{equation}\label{2.2}
\hat{p}_{\mu}=\pi_{\mu}+\frac{\epsilon}{2}\left( \left(-\frac{1}{8} u + \frac{3}{8} v+ \frac{1}{4}z\right)\pi_{\mu}+\frac{M}{R} \xi_{\mu}\left( -\frac{3}{8} u + \frac{1}{8} v- \frac{1}{4}z\right) +h.c. \right),
\end{equation}
\begin{equation}
\hat{h}=1-\frac{\epsilon}{2}\left( u-v\right), \quad M_{\mu\nu}=\xi_{\mu}\pi_{\nu}-\xi_{\nu}\pi_{\mu},
\end{equation}
where
\begin{equation}
u=\frac{1}{M^{2}}\pi^{2}, \quad v=\frac{1}{R^{2}}\xi^{2}, \quad z=\frac{1}{MR}\xi\cdot\pi,\quad
(u)^{\dagger}=u, \quad (v)^{\dagger}=v, \quad (z)^{\dagger} \neq z,
\end{equation}

From \eqref{2.1} and \eqref{2.2} it follows that
\begin{equation}
	M\left( \hat{x}_{\mu}-\xi_{\mu}\right) =R\left( \hat{p}_{\mu}-\pi_{\mu}\right).
\end{equation}

\section{Conformal algebra}
In this section we summarize the main properties of the conformal algebra, for comparison with the previous results. This is the symmetry algebra of Minkowski spacetime, and is
generated by Lorentz generators $M_{\mu\nu}$, generators of translations $P_{\mu}$, the dilatation generator $D$ and conformal generators $K_{\mu}$ obeying the following commutations relations
\begin{equation}\label{c1}
\left[ D, K_{\mu}\right] =-iK_{\mu},
\end{equation}
\begin{equation}\label{c2}
\left[ D, P_{\mu}\right] =iP_{\mu},
\end{equation}
\begin{equation}\label{c3}
	\left[ K_{\mu}, P_{\nu}\right] =2i\left( \eta_{\mu\nu}D+M_{\mu\nu}\right) ,
\end{equation}
\begin{equation}\label{c4}
\left[ P_{\mu}, P_{\nu}\right] =\left[ K_{\mu}, K_{\nu}\right] =0,
\end{equation}
giving rise to an $o(2,4)$ algebra.
Choosing $MR=1, \epsilon =-1$ and making the substitutions $\tilde{X}_{\mu} \rightarrow \frac{1}{2}K_{\mu},\;  \tilde{P}_{\mu} \rightarrow P_{\mu},\; \tilde{H} \rightarrow D $ in \eqref{1.8}-\eqref{1.12}, we observe that the generalized Heisenberg algebra \eqref{1.8}-\eqref{1.12} becomes the conformal algebra  \eqref{c1}-\eqref{c4}.

The standard Hermitian realizations of the conformal algebra on phase space is
\begin{align}
&M_{\mu\nu}=x_{\mu}p_{\nu}-x_{\nu}p_{\mu}, \quad D=\frac{1}{2}\left( x\cdot p+p\cdot x\right) , \quad P_{\mu}=p_{\mu},\\
&K_{\mu}=x_{\mu}x\cdot p-\frac{1}{2}x^{2}p_{\mu}+h.c.,
\end{align}
where $x$ and $p$ are canonical variables satisfying
\eqref{2}.

The relations between the generators of the generalized Heisenberg algebra \eqref{1.15}-\eqref{1.18} and the conformal algebra are given by
\begin{equation}
	\tilde{P}_{\mu}=P_{\mu}=p_{\mu}, \quad \tilde{X}_{\mu}=x_{\mu}-\frac{\epsilon}{2MR}K_{\mu}, \quad \tilde{H}=1-\frac{\epsilon}{MR}D,
\end{equation}
while the relations between the generators of the Yang model and the conformal algebra are
\begin{equation}
\hat{x}_{\mu}=\xi_{\mu}-\frac{\epsilon \sqrt{2}}{4MR}K_{\mu}, \quad \hat{p}_{\mu}=\pi_{\mu}-\frac{\epsilon \sqrt{2}}{4R^{2}}K_{\mu}, \quad
\hat{h}=1-\frac{\epsilon}{MR}D.
\end{equation}

\section{Final remarks}
The $o(2,4)$ algebra is at the basis of several different physical models, like the conformal symmetries of Minkowski space and some variants of the Yang model
of noncommutative geometry on curved spacetime. In this letter we have proposed still another application of  $o(2,4)$, defining a generalization of the Heisenberg algebra related to a deformation of the quantum phase space, with an operator-valued Planck constant.

We have discussed some algebraic relations between these models, and outlined some implications of the generalized Heisenberg algebra, but a more thorough investigation of these would be worthy, in particular the details of  their associated Hopf algebra. We also notice that these models give rise to a deformation of the standard uncertainty principle, as in similar cases \cite{MM-3,Mignemi-2}: it would be interesting to check whether they also imply the existence of minimal lengths or momenta, as in simpler cases \cite{MM-3}.
We plan to discuss these topics in a future publication.

\section*{Acknowledgements}
S. Mignemi would like to thank Partha Nandi for an illuminating discussion and  to acknowledge the contribution of GNFM and of COST Action CA23130.

This research was partially funded by the European Union (NextGenerationEU) under the Croatian Recovery and Resilience Plan 2021–2026 (NRRP), through the University of Split institutional project “Mathematical modeling and simulations of physical systems (MaMoS) IP-UNIST-46”, approved by the Ministry of Science, Education and Youth of the Republic of Croatia.

The views and opinions expressed are solely those of the author(s) and do not necessarily reflect those of the European Union or the European Commission. Neither the European Union nor the European Commission can be held responsible for them.


\begin{thebibliography}{99}

\bibitem{Ka2} H.A. Kastrup, Position noperators, gauge transformations and the conformal group, \PR{143}, 1021 (1966).

\bibitem{Na} P. Nandi and F.G. Scholz, The hidden Lorentz covariance of quantum mechanics, \AoP{404}, 169643 (2024).

\bibitem{Ka} H.A. Kastrup, Conformal group and its connection with an indefinite metric in Hilbert space, \PR{140}, B183 (1965).

 H.A. Kastrup, Conformal group in spacetime, \PR{142}, 1060 (1966).

\bibitem{bar}A.O. Barut and and W.E. Brittin, ed., De Sitter and conformal groups and their applications, Lectures in Theoretical Physics \textbf {13}, Colorado Associated University Press, Boulder, 1971.

\bibitem{Ya} C.N. Yang, {On Quantized Space-Time}, \PR{72}, 874 (1947).

\bibitem {chr} C. Chryssomalakos and E. Okon, Generalized quantum relativistic kinematics: a stability point of view, \emph{Int. J. Mod. Phys.} \textbf{D 13}, 1817 (2004).

\bibitem{guo} H.G. Guo, C.G. Huang and H.T. Wu, Yang's model as Triply Special Relativity and the Snyder's model-de Sitter special relativity duality, \emph{Phys. Lett.} \textbf{B 663}, 270 (2008).

\bibitem{ver}J.  Heckman and H. Verlinde, Covariant non-commutative spacetime, \NP{ B 894}, 58 (2015).

\bibitem{luk} J. Lukierski and M. Woronowicz, Spinorial Snyder and Yang models from superalgebras and noncommutative quantum superspaces,   \PL{B 824}, 136783 (2022).

\bibitem{MM-2}
 S. Meljanac and S. Mignemi, {Generalizations of Snyder model to curved spaces}, \PL{B 833}, 137289 (2022).

S. Meljanac, S. Mignemi, {Noncommutative Yang model and its generalizations}, {\JMP{64}},  {023505 (2023)}.

\bibitem{Lukierski}
\bibp{ J. Lukierski, S.  Meljanac, S.  Mignemi and A. Pacho\l{}}{Quantum pertubative solutions of extended Snyder and Yang models with spontaneous symmetry breaking}{\PL{B 847}}, {138261 (2023)}.

 \bibp{J. Lukierski, S. Meljanac, S. Mignemi, A. Pacho\l\ and M. Woronowicz} {From Snyder space-times to doubly $\kappa$-dependent Yang quantum phase spaces and their generalizations} {\PL{B 854}, 138729 (2024)}.

\bibitem{Martinic}
\bibp{T. Martini\'{c} Bila\'{c}, S. Meljanac, S. Mignemi}{Hermitian realizatons of the Yang model}{\JMP{64}}, 122302 (2023).

 T. Martini\'c-Bila\'c,  S. Meljanac and S. Mignemi, {Realizations and star-product of doubly $\kappa$-deformed Yang models,} \EPJ{C 84}, 846 (2024).

T. Martini\'c-Bila\'c,  S. Meljanac and S. Mignem, {Dual $\kappa$-Minkowski spaces from Yang model and their Weyl realizations}, \JMP{66}, 092301 (2025).

\bibitem{MM}
\bibp{S. Meljanac, S. Mignemi}{Reduced Yang model and noncommutative geometry of curved spacetime} \MPL{A 40}, 37, 2550187 (2025).


\bibitem{MM-3} S. Meljanac, S. Mignemi, {Quantum mechanics of the nonrelativistic Yang model}, {\EL{150}, 39001 (2025)}.


\bibitem{ste} A. Manta and H.C. Steinacker,  Minimal covariant quantum space-time,  \JoP{A 58}, 175204  (2025).


\bibitem{pac} A. Pacho\l, Generalized Extended Uncertainty Principles, Liouville theorem and density of states: Snyder-de Sitter and Yang models, \NP{B 1010}, 116771 (2025)

\bibitem{kup} V.G. Kupriyanov and E.L.E. de Lima, {Symplectic realization of generalized Snyder-Poisson algebra}, \emph{Universe} \textbf{11}, 339 (2025).

\bibitem{CG}
G.P. Mangano, F. Lizzi and A. Porzio, Inconstant Planck's constant, \IJMP{A30}, 1550209 (2015).

J.L. Cort\'es and J. Gamboa, Deformed classical-quantum mechanics transitions, \PR{D102}, 036015 (2020).

\bibitem{Mignemi-2}
\bibp{S. Mignemi}{Classical and quantum mechanics of the nonrelativistic Snyder model in curved space}{\textit{Class. Quantum Grav.}}{ \textbf{29}, 215019 (2012)}.


\end{thebibliography}
\end{document}